\documentclass[11pt]{article}
\usepackage{longtable,supertabular}
\usepackage{amsmath}
\usepackage{amssymb}
\usepackage{latexsym}

\title{\bf GHW-based Necessary and Sufficient Conditions for Support Constrained Generator Matrices}
\author{Hao Chen
  \thanks{Hao Chen is with the College of Information Science and Technology/Cyber Security, Jinan University, Guangzhou, Guangdong Province, 510632, China, haochen@jnu.edu.cn. The research of Hao Chen was supported by NSFC Grant 62032009.}}

\begin{document}

\maketitle
\begin{abstract}
Support constrained generator matrix for a linear code has been an active topic in recent years. The necessary and sufficient condition for the existence of MDS codes over small fields with support constrained generator matrices were conjectured by Dau, Song, Yuen in 2014. This GM-MDS conjecture was proved independently by Lovett and Yildiz-Hassibi in 2018. In this paper we propose the necessary and sufficient conditions for support constrained generator matrices of general linear codes based on the generalized Hamming weights. It is proved that  the direct generalization of the GM-MDS conjecture for $2$-MDS codes and algebraic geometry codes is not true over arbitrary fields. We propose and prove a GHW-based sufficient condition for support constrained matrices of general linear codes. This is the first sufficient condition for the existence of support constrained generator matrices for general linear codes over arbitrary finite fields. Moreover a weaker GHW-based sufficient condition for support constrained generator matrices are given for the binary simplex code and the first order $q$-ary Reed-Muller codes.\\
\end{abstract}

\section{Introduction}

For a vector ${\bf a} \in {\bf F}_q^n$, the Hamming weight $wt({\bf a})$ of ${\bf a}$ is the number of non-zero coordinate positions. The Hamming distance $d_1({\bf a}, {\bf b})$ between two vectors ${\bf a}$ and ${\bf b}$ is defined to be the Hamming weight of ${\bf a}-{\bf b}$. A linear $[n,k]_q$ code is a $k$ dimensional subspace ${\bf C} \subset {\bf F}_q$. For a linear code ${\bf C} \subset {\bf F}_q^n$ of dimension $k$, its Hamming distance (or weight) $d_1$ is the minimum of Hamming distances $d_1({\bf a}, {\bf b})$ between any two different codewords ${\bf a}$ and ${\bf b}$ in ${\bf C}$. It is well-known that the Hamming distance (or weight) of a linear code ${\bf C}$ is the minimum Hamming weight of non-zero codewords. These codewords of a linear code ${\bf C}$ with the minimum Hamming weight are called minimum weight codewords.\\

The Singleton bound $d_1 \leq n-k+1$ is the basic upper bound for linear error-correcting codes. A linear code attaining this bound is called a MDS (maximal distance separable) code. A long-standing conjecture in the theory of linear error-correcting codes is the main conjecture for the MDS codes, which asserts that the length of a linear MDS code over ${\bf F}_q$ can not bigger than $q+1$ except some obvious trivial cases, we refer to page 264-265 of \cite{HP}. One achievement is the work \cite{Ball}  of Ball in which the conjecture was proved for linear MDS codes over prime fields.\\

Let ${\bf F}_q$ be an arbitrary finite field, $P_1,\ldots,P_n$ be $n \leq q$ elements in ${\bf F}_q$. The Reed-Solomon code $RS(n,k)$ is defined by $$RS(n,k)=\{(f(P_1),\ldots,f(P_n)): f \in {\bf F}_q[x],\deg(f) \leq k-1\}.$$ This is an $[n,k,n-k+1]_q$ linear MDS code, since a degree $\deg(f) \leq k-1$ polynomial has at most $k-1$ roots. From the interpolation it is clear that for any given $k-1$ distinct points among $P_1,\ldots,P_n$ there is one nonzero polynomial $f$ of degree $$\deg(f) \leq k-1$$ such that $f$ is zero at these $k-1$ points. That is, every subset of $\{P_1,\ldots,P_n\}$  with the cardinality $k-1$ is the the set of zero coordinate positions of at least one nonzero codeword.\\

The Reed-Solomon codes can be generalized to algebraic-geometric codes as follows. Let ${\bf X}$ be an absolutely irreducible  non-singular genus $g$ curve defined over ${\bf F}_q$. Let ${\bf P}=\{P_1,\ldots,P_n\}$ be a set $n$ distinct rational points of ${\bf X}$ over ${\bf F}_q$. Let ${\bf G}$ be a rational divisor over ${\bf F}_q$ of degree $\deg({\bf G})$ satisfying $2g-2 <\deg({\bf G})<n$  and $$support({\bf G}) \bigcap {\bf P}=\emptyset.$$
 Let  ${\bf L}({\bf G})$ be the associated function space. The algebraic-geometric code associated with ${\bf G}$, $P_1,\ldots,P_n$ is defined by $${\bf C}({\bf P}, {\bf G}, {\bf X})=\{(f(P_1),\ldots,f(P_n)): f \in {\bf L}({\bf G})\}.$$ From the Riemann-Roch Theorem the dimension of this code is $$k=\deg({\bf G})-g+1.$$  The minimum Hamming distance is $$d_1 \geq n-\deg({\bf G}).$$ The Reed-Solomon codes are just the algebraic-geometric codes over the genus $0$ curve. As in the case of the Reed-Solomon codes, from the Riemann-Roch Theorem any subset of $\{P_1,\ldots,P_n\}$ of cardinality $\deg({\bf G})-g$ is in the set of zero coordinate positions of at least one nonzero codeword.One achievement of the theory of algebraic-geometric codes is the sequence of algebraic-geometric codes over ${\bf F}_{q^2}$ satisfying $$R+\delta \geq 1-\frac{1}{q-1},$$ which is exceeding the Gilbert-Varshamov bound.  We refer to \cite{TV} for the detail.\\

The $q$-ary Reed-Muller codes are defined as follows. Let $P_1,\ldots,P_n$ be $n=q^m$ points of ${\bf F}_q^m$. Let $u\leq m$ be a positive integer. Let ${\bf Function}(u,m)$ be the set of linear combinations of monomials $x_{i_1}^{j_1} x_{i_2}^{j_2} \cdots x_{i_t}^{j_t}$, where $j_i \leq q-1$ for $i=1,\ldots,t$, $t \leq u$.  The $u$ order Reed-Muller code $RM(u,m)$ is defined by $$RM(u,m)=\{(f(P_1),\ldots,f(P_n)):f \in {\bf Function}(u,m)\}.$$ We consider the first order $q$-ary Reed-Muller code $RM(1,m)$. The dimension is $$k=1+m,$$  and the minimum distance is $$d_1=(q-1)q^{m-1}.$$\\

The binary simplex code is an $[2^m-1, m, 2^{m-1}]_2$ linear code with one nonzero weight $2^{m-1}$, see \cite{HP}. For any given linear function $a_1x_1+\cdots+a_mx_m$, the codeword is the evaluation vector of this linear function at $2^m-1$ nonzero vectors of $({\bf F}_2^n)^*$. Hence the only nonzero weight is $2^{m-1}$.\\

The generalized Hamming weights (GHW) of linear codes was introduced in \cite{HKM} in 1977, see page 283 of \cite{HP},  and studied further in \cite{Wei}. These parameters are naturally related to the support constrained matrices. The support of a linear sub-code $D \subset {\bf C}$ is $$\operatorname{supp}(D)=\{1 \leq i \leq n: x_i \neq 0: \exists x=(x_1,\ldots,x_n) \in D\},$$ that is, the support of a linear sub-code $D$ is the non-zero coordinate positions of all codewords in $D$. The $r$-th generalized Hamming weight $d_r$ for $1\leq r \leq k$ is defined to be the minimum of the number of support positions of arbitrary $r$ dimension sub-codes. Hence $d_1$ is the minimum Hamming weight. It is clear that $d_1<d_2 < \cdots <d_k$ and the generalized Singleton bound $d_r \leq n-k+r$ is satisfied for a linear $[n,k]_q$ code. Moreover the $r$-th generalized Hamming weight of a linear $[n,k]_q$ MDS code is exactly $n-k+r$ for $1 \leq r \leq k$. We refer to \cite{Wei} for the other properties of the generalized Hamming weights. The weight hierarchy or generalized Hamming weights of many well-constructed linear codes, such as cyclic codes, Hermitian codes, $q$-ary Reed-Muller codes have been determined in \cite{GV,GV1,Munuera}.\\

Let ${\bf G}$ be a $k \times n$ generator matrix of a linear $[n,k,n-k+1]_q$ MDS codes, set ${\bf S}_i \subset [n]=\{1,2,\ldots,n\}$ be the set of zero coordinate positions of the $i$-th row of this generator matrix ${\bf G}$. Then it follows from the MDS condition that for any subset $I \subset [k]=\{1,2,\ldots,k\}$, the intersection $\bigcap_{i \in I} {\bf S}_i$ has at most $k-|I|$ elements. The GM-MDS conjecture claims that for given $n$ and $k$, and any given subset system ${\bf S}_1,\ldots, {\bf S}_k$ of $[n]$ satisfying the MDS condition  $$|I|+|\bigcap_{i\in I} {\bf S}_i| \leq k,$$ if $$q \geq n+k-1,$$ then there is a $k\times n$ generator matrix ${\bf G}$ of a linear $[n,k,n-k+1]_q$ MDS code over ${\bf F}_q$ such that ${\bf S}_i$ is in  the set of zero coordinate positions of the $i$-th row of ${\bf G}$ for $i=1,2,\ldots,k$.\\

The above conjecture was formulated by Dau, Song and Yuen in \cite{DSY14} from the motivation of multiple access network coding scheme. It was found its application in weakly secure document exchange in \cite{YSZ14}. There have been a lot of progress in \cite{DSY15,HLH,HLH1,HS,SC} since this conjecture was formulated in \cite{DSY14}. This GM-MDS conjecture was finally proved in two independent papers \cite{YH2,Lovett}. In \cite{GS19} further results about smaller field with size $n$ was proved under stronger conditions on the subset systems. In \cite{YH4} Gabidulin codes with support constrained generator matrices were constructed. For other development and related conjectures we refer to \cite{KLR,Lovett,GGY}. In particular it was conjectured that for general support constraints generator matrices can only be existent over exponential size fields in \cite{Lovett}. This kind general problem is about random codes and thus theoretically significant. \\

In this paper we first prove in Theorem 2.1 that the MDS condition in the GM-MDS conjecture is a special case of a general necessary condition for support constrained generator matrices of general linear codes based on their generalized Hamming weights. Then it is natural to consider the following problem. For a general linear code over ${\bf F}_q$, what support set conditions should be imposed for generator matrices? Can a  generator matrix with support constraints satisfying the support set conditions in Theorem 2.1 be explicitly given? This problem is motivated from the proof  of the GM-MDS conjecture. Theoretically this is also a combinatorial problem for the structures of linear codes. The direct generalization of Gm-MDS conjecture to $2$-MDS codes and algebraic geometry codes is proved not true in Section 3. The GHW-based sufficient conditions for support constrained generator matrices of general linear codes are given in Theorem 2.2 and Corollary 2.1. In Section 4 a weaker sufficient condition for support constrained generator matrices of the binary simplex code and the first order $q$-ary Reed-Muller codes are also presented.\\

\section{The GHW-based necessary and sufficient conditions for support constrained matrices}

We prove the following result, from which the MDS condition in the GM-MDS conjecture is a special case.\\

{\bf Theorem 2.1.} {\em Let ${\bf C}$ be a linear $[n, k]_q$ code over the finite field ${\bf F}_q$ with the generalized Hamming weights $d_1<d_2 <\cdots<d_k$. Let ${\bf G}$ be a size $k \times n$ generator matrix of this code ${\bf C}$. Set ${\bf S}_i$, $i=1,2,\ldots,k$, the set of zero coordinate positions of the $i$-th row of this generator matrix ${\bf G}$. Then we have $$|\bigcap_{i \in I} {\bf S}_i| \leq n-d_{|I|}$$ for any subset $I \subset [k].$}\\

{\bf Proof.} For any subset of size $I$ in the set $[k]=\{1,2,\ldots,k\}$, the $|I|$ corresponding rows of ${\bf G}$ generate a $|I|$ dimensional sub-code of ${\bf C}$, hence we have $$d_{|I|} \leq n- |\bigcap_{i \in I} {\bf S}_i|.$$ The conclusion follows immediately.\\

For a linear $[n,k]_q$ MDS code, we have $d_r=n-k+r$. Then we have $|I|+|\bigcap_{i \in I} {\bf S}_i| \leq k$. This is the MDS condition on the set systems in the GM-MDS conjecture. Hence the GM-MDS conjecture due to Dau, Song, Yuen and Hassibi in \cite{DSY14,HHYD} can be reformulated as follows. If the subset system ${\bf S}_1,\ldots,{\bf S}_k$ in $[n]$ satisfies the generalized Hamming weight based constraints for a linear $[n,k]_q$ MDS code, $$|\bigcap_{i \in I} {\bf S}_i| \leq n-d_{|I|},$$ where $d_1,\ldots, d_k$ are the generalized Hamming weights of a linear $[n,k]_q$ MDS code,  if $q \geq n+k-1$, is there a linear $[n,k]_q$ MDS code with a support constrained generator matrix ${\bf G}$ such that the set ${\bf S}_i$ is in the set of zero coordinate positions of the $i$-th row of ${\bf G}$ for $i=1,2,\ldots,k$.\\

A generalized MDS condition for the MDS condition in the GM-MDS conjecture was proposed in \cite{Lovett} and studied in \cite{GS19}.  Our reformulation of the MDS condition based on the generalized Hamming weights in the GM-MDS conjecture is for linear codes and motivated from the existence of support constrained generator matrices. We can consider the following problem. Let ${\bf C}$ be a linear $[n, k]_q$ code over the finite field ${\bf F}_q$ with the generalized Hamming weights $d_1<d_2 <\cdots<d_k$. Let ${\bf S}_1, \ldots, {\bf S}_k$ be a subset system in $[n]$ satisfying $$|\bigcap_{i \in I} {\bf S}_i| \leq n-d_{|I|}.$$ Is there a corresponding support constrained generator matrix ${\bf G}$ such that for $i=1,2,\ldots,k$, the set ${\bf S}_i$ is in the set of zero coordinate positions of the $i$-th row of ${\bf G}$? The answer to this problem is negative as showed in Section 3.\\

From Theorem 2.1 the  GM-MDS conjecture can be considered in coding-dependent version. That is, the conjecture is only considered for the generalized Hamming weights for one concrete code, not a class of codes.  The following GHW-based sufficient condition in Corollary 2.1 for support constrained generator matrices seems plausible.  Notice that the condition 1) is satisfied automatically for any subset of cardinality $k-1$ for linear  $[n,k,n-k+1]_q$ MDS codes, thus the following two results are natural extensions of the GM-MDS conjecture to general linear codes.\\

{\bf Theorem 2.2.}  {\em Let ${\bf C}$ be a linear $[n, k]_q$ code over the finite field ${\bf F}_q$. Let ${\bf S}_i$, $i=1,2,\ldots,k$, be a subset system in $[n]$ satisfying the following two conditions.\\
1)  Each ${\bf S}_i$ is the set of zero coordinate positions of a nonzero codeword of ${\bf C}$,\\
2) For any two subsets $I_1\subset I_2 \subset [k]$ satisfying $I_1 \neq I_2$, $|\bigcap_{i \in I_2} {\bf S}_i| < |\bigcap_{i \in I_1} {\bf S}_i|.$\\
 Then there is a support constrained generator matrix generator matrix ${\bf G}$ such that for $i=1,2,\ldots, k$, the set ${\bf S}_i$ is the set of zero coordinate positions of the $i$-th row of ${\bf G}$.}\\

 {\bf Proof.} Let ${\bf c}_i$ be the codeword of ${\bf C}$ such that ${\bf S}_i$ is its set of zero coordinate positions. We only need to prove these codewords are linearly independent.  If ${\bf c}_{i_1}, \ldots, {\bf c}_{i_t}$ are linear dependent, then $${\bf c}_{i_u}=\Sigma_{j \in [t]-u} a_j{\bf c}_{i_j}, $$ where $a_j \in {\bf F}_q$. Therefore ${\bf S}_{i_u} \subset \bigcap_{j \in [t]-u} {\bf S}_{i_j}$. Set $I_2=\{i_1,\ldots,i_t\}$ and $I_1=I_1 -\{i_u\}$, then $$\bigcap_{j \in [t]} {\bf S}_{i_j}=\bigcap_{j \in [t]-u} {\bf S}_{i_j}.$$ The second condition is not satisfied. The conclusion follows.\\

It is clear that the above sufficient condition is not necessary as the follow generator matrix of the linear $[4,3,2]_3$ ternary code.\\

$$
\left(
\begin{array}{cccccc}
1&1&0&0\\
0&0&1&1\\
1&-1&1&-1\\
\end{array}
\right)
$$
Let ${\bf S}_i$, $i=1,2,3$ be the zero coordinate positions of the $i$-th row. Then ${\bf S}_1=\{3,4\}$, ${\bf S}_2=\{1,2\}$ and ${\bf S}_3=\emptyset$. The three codewords are linear independent and this is the generator matrix. However $|{\bf S}_1 \bigcap {\bf S}_2 \bigcap {\bf S}_3|=|{\bf S}_1 \bigcap {\bf S}_2|$.\\

The following GHW-based sufficient condition for support constrained generator matrix follows from Theorem 2.1 immediately.\\

{\bf Corollary 2.1 (The GHW-based sufficient condition for support constrained generator matrices).} {\em Let ${\bf C}$ be a linear $[n, k]_q$ code over the finite field ${\bf F}_q$ with the generalized Hamming weights $d_1<d_2 <\cdots<d_k$. Let ${\bf S}_1, \ldots, {\bf S}_k$ be a subset system in $[n]$ satisfying the following two conditions.\\
1)  ${\bf S}_i$ is the set of zero coordinate positions of a nonzero minimum weight codeword of ${\bf C}$,\\
2) For any subset $I \subset [n]$, $|\bigcap_{i \in I} {\bf S}_i| =n-d_{|I|}.$\\
Then there is a support constrained generator matrix generator matrix ${\bf G}$ such that for $i=1,2,\ldots, k$, the set ${\bf S}_i$ is the set of zero coordinate positions of the $i$-th row of ${\bf G}$.}\\

It is obvious that linear codes in Corollary 2.1 are generated by minimum weight codewords. If a linear code is generated by its minimum weight codewords, then the above GHW-based sufficient condition is suitable to determine which subset systems can be zero positions of rows of generator matrices. This class of linear codes are quite common in coding theory. For example, Reed-Solomon codes, some algebraic geometry codes, Reed-Muller codes and many linear codes with few nonzero weights as in \cite{Ding,Ding1,Ding2}, are generated by minimum weight codewords.\\

\section{The direct generalization of the GM-MDS conjecture for $2$-MDS codes is not true}

When the generalized Hamming weights are from the linear $[n,k]_q$ MDS code, the necessary condition in Theorem 2.1 is sufficient for MDS codes over small fields, as proved independently in \cite{Lovett,YH2}. In \cite{GS19} the smaller field size results are proved if some further stronger restraints on the subset systems are imposed.  Algebraic geometry codes over elliptic curves are natural generalizations of Reed-Solomon codes. Hence it is interesting to consider the possible generalization of GM-MDS conjecture and then a beautiful theorem to algebraic geometry codes over elliptic curves. Theorem 2.2 and Corollary 2.1 are natural extensions in this case, however the sufficient conditions in Theorem 2.2 and Corollary 2.1 are clearly much stronger than the necessary.\\

A linear $[n, k]_q$ code over ${\bf F}_q$ is called $r$-MDS for some $r$ in the range $1\leq r \leq k$, if $d_r=n-k+r$. Then it is also $s$-MDS for any $s \geq r$, see \cite{TV1}. The linear MDS codes are then $1$-MDS.  Hence $r$-MDS codes for $r\ge 2$ are natural generalizations of linear MDS codes.  A well-known result in weight hierarchy or higher weights about algebraic-geometric codes due to Tsfasman and Vl\v{a}dut is that these codes are $g+1$-MDS if they are from genus $g$ curves, see \cite{TV1} Corollary 4.2. As algebraic-geometric codes from genus $0$ curves, the Reed-Solomon codes are MDS ($1$-MDS). The next interesting cases are these algebraic-geometric $2$-MDS codes from elliptic curves.\\

Since the GM-MDS conjecture are about $1$-MDS linear codes, we can consider the direct generalization of the GM-MDS conjecture for $2$-MDS linear codes.  The  generalized Hamming weights of $2$-MDS linear (not MDS) codes are as follows, $$d_1=n-k,$$ $$d_2=n-k+2,$$ $$\cdots,$$ $$d_r=n-k+r,$$ $$\cdots,$$ $$d_k=n.$$ Many algebraic-geometric $[n,k]_q$ codes from elliptic curves with code lenght $n>q+2$ have their generalized Hamming weights as above. However for algebraic-geometric code from elliptic curve cases, not every subset of $[n]$ of the cardinality $k$ can be the set of zero coordinate positions of nonzero codeword, the condition $|{\bf S}_i| \leq k-1$ is a natural constraint.\\

Therefore the GHW -based support constrained conditions on the subset systems for two or more subsets are the same as the MDS condition in the GM-MDS conjecture. The case of one subset constraint $|{\bf S}_i|\leq k-1$ is stronger than the condition $$|{\bf S}_i|  \leq n-d_1=k.$$  \\

{\bf The direct generalization of the GM-MDS conjecture for $2$-MDS codes.} {\em. Let ${\bf S}_1, \ldots, {\bf S}_k$ be a subset system in $[n]$ with their cardinalities smaller than or equal to $k-1$. If this subset system satisfies the  constraints $$|\bigcap_{i \in I} {\bf S}_i|+|I| \leq k$$ for any subset $I \subset [k]$ with $|I|\geq 2$. Over sufficiently large field ${\bf F}_q$ we have a linear $[n,k,n-k]_q$ $2$-MDS code and a support constrained generator matrix ${\bf G}$ such that  for $i=1,2,\ldots, k$, the set ${\bf S}_i$ is in the set of zero coordinate positions of the $i$-th row of ${\bf G}$.}\\

Let ${\bf C}$ be the elliptic curve over ${\bf F}_4$ defined by $y^3=x^2+x$ over ${\bf F}_4$. Let $\omega$ be the element in ${\bf F}_4$ such that $\omega^2+\omega+1=0$. Then the $8$ rational points of the above elliptic curve is of the form $P_1=(0,0),P_2=(1,0),P_3=(\omega,1),P_4=(\omega,\omega),P_5=(\omega,\omega^2),P_6=(\omega^2,1),P_7=(\omega^2,\omega),P_8=(\omega^2,\omega^2)$. We have a dimension $3$ algebraic-geometric $[8,3,5]_4$ code from this elliptic curve with the following generator matrix. The 2nd and 3rd generalized Hamming weights are $d_2=7$ and $d_3=8$.

$$
\left(
\begin{array}{cccccccc}
1&1&1&1&1&1&1&1\\
0&1&\omega&\omega&\omega&\omega^2&\omega^2&\omega^2\\
0&0&1&\omega&\omega^2&1&\omega&\omega^2\\
\end{array}
\right)
$$

Set ${\bf S}_1=\{P_4,P_8\}$, ${\bf S}_2=\{P_3,P_7\}$ and ${\bf S}_3=\{P_5,P_6\}$. Then every set has $k-1=2$ elements.  Moreover this subset system satisfies the GHW-based constraint in the GM-$2$-MDS conjecture. For $i=1,2,3$, the functions of the form  $a_0+a_1x+a_2y$, $a_i \in {\bf F}_4$, for $i=1,2,3$, which are zero at the two points in ${\bf S}_i$ have to be the form $a(x+\omega^{i-1} y)$ for an arbitrary nonzero element $a \in {\bf F}_4$. Hence these functions are zero at the point $P_1$. It is clear that the three functions of the above forms are linearly dependent. Hence there is no $3 \times 8$ generator matrix ${\bf G}$ of this algebraic-geometric code such that ${\bf S}_1, {\bf S}_2, {\bf S}_3$ are in the zero coordinate position sets of rows in ${\bf G}$. This is a counter-example for the  GM-$2$-MDS conjecture.\\

Let $P^2({\bf F}_q)$ be the projective plane over the finite field ${\bf F}_q$, ${\bf E} \subset {\bf P}^2({\bf F}_q)$ be an absolutely non-singular degree $3$ curve. Then the genus of this curve is $\frac{(3-2)(3-1)}{2}=1$ and ${\bf E}$ is an elliptic curve. Let ${\bf G}$ be a degree $3$ rational divisor defined by a line $a_0x_0+a_1x_1+a_2x_2=0$, where $x_0,x_1,x_2$ are homogeneous coordinates of ${\bf P}^2({\bf F}_q)$, $a_0, a_1, a_2 \in {\bf F}_2$.  Let ${\bf P}$ be a set of $n$ rational  points of ${\bf E}$.  We assume $support({\bf G}) \bigcap {\bf P}=\emptyset$. Then ${\bf C}({\bf P}, {\bf G}, {\bf E})$ is a dimension $3$ algebraic-geometric code from ${\bf E}$. Assume that there are three points $P_1, P_2, P_3$ in ${\bf P}$ which are on one line, then the minimum Hamming weight of this code is $n-3$. Hence this is a linear $[n, 3, n-3]_q$ code which is also $2$-MDS.\\

{\bf Theorem 3.1.} {\em The direct generalization of the GM-MDS conjecture for $2$-MDS codes is not true for above $3$-dimensional algebraic-geometric codes ${\bf C}({\bf P}, {\bf G}, {\bf E})$  from degree $3$ plane curves.}\\

{\bf Proof.} Since the functions in ${\bf L}({\bf G})$ are of the form $$\frac{b_0x_0+b_1x_1+b_2x_2}{a_0x_0+a_1x_1+a_2x_2},$$ where $b_0, b_1, b_2 \in {\bf F}_q$. Then from the condition 1 in the code-depending version of SCGM-GHW conjecture the subset ${\bf S}_i \subset {\bf P}$, $i=1,2,3$, is  consisting of three points in ${\bf P}$ which are on one projective line. Then condition ${\bf S}_i \bigcap {\bf S}_j$ has at most one point. Thus the GHW-based constraints for $|I|=2$ is satisfied automatically. The condition ${\bf S}_1 \bigcap {\bf S}_2 \bigcap {\bf S}_3$ is empty is valid for three linear independent lines passing through one point outside this elliptic curve. Hence these three projective lines are linearly  dependent. The conclusion follows directly.\\

\section{Support constrained generator matrices for the binary simplex code and the first order $q$-ary Reed-Muller code}

The main result in this Section is the following sufficient condition for support constrained generator matrices of the binary simplex code and the first order Reed-Muller codes.\\

Let ${\bf C}$ be a linear $[n, k]_q$ code over the finite field ${\bf F}_q$ with the generalized Hamming weights $d_1<d_2 <\cdots<d_k$. Let ${\bf S}_1, \ldots, {\bf S}_k$ be a subset system in $[n]$ satisfying the following two conditions.\\
1) $|{\bf S}_i|=n-d_1$ and for each ${\bf S}_i$ there is a nonzero codeword of weight $d_1$ such that ${\bf S}_i$ is its set of zero coordinate positions,\\
2) This subset system satisfies the GHW-based constraints $$|\bigcap_{i \in I} {\bf S}_i| \leq n-d_{|I|}.$$
Then is there a support constrained generator matrix generator matrix ${\bf G}$ such that for $i=1,2,\ldots, k$, the set ${\bf S}_i$ is the set of zero coordinate positions of the $i$-th row of ${\bf G}$?\\

The generalized Hamming weights of the first order $q$-ary Reed-Muller code $RM(1,m)$ are as follows, see \cite{HP98}, $$d_1=(q-1)q^{m-1},$$ $$d_2=q^m-q^{m-2},$$ $$d_r=q^m-q^{m-r},$$ $$d_m=q^m-1,$$ and $$d_{m+1}=q^m.$$ The generalized Hamming weights for binary simplex $[2^m-1,m,2^{m-1}]_2$ code are calculated in \cite{Wei} as follows. $$d_r=2^{m-1}+\cdots+2^{m-r}, $$ we refer to Corollary 3 in \cite{Wei}.\\

{\bf Theorem  4.1.} {\em The above GHW-based conditions are sufficient for support constrained generator matrices for the first order $q$-ary Reed-Muller codes and the binary simplex code}\\

{\bf Proof.} Each ${\bf S}_i$ is the zero subspace of an affine function  of the form $$a_0+a_1x_1+\cdots+a_mx_m,$$ where $a_i \in {\bf F}_q$ for $i=0,1,\ldots,m$. Hence there are exactly $q^{m-1}$ points in ${\bf S}_i$ for $i=1,2,\ldots,m+1$. From the GHW-based constraints $$|\bigcap_{i \in I} {\bf S}_i|\leq q^m-d_{|I|},$$ $|I|$ such affine spaces intersect at a dimension $m-|I|$ affine subspace, when $|I| \leq m$, and intersect at the empty set when $|I|=m+1$. Hence we can find exactly $m+1$ linearly independent affine functions such ${\bf S}_i$, $i=1,2,\ldots,m+1$ is exactly the set of zero coordinate positions of the $i$-th affine function. The conclusion for binary simplex code can be proved similarly.\\

It would be interesting to consider for what linear codes the above GHW-based condition is sufficient for the existence of support generator matrices. The conditions is weaker than the sufficient condition proved in Corollary 2.1. Though algebraic geometry codes over elliptic curves are natural generalizations of Reed-Solomon code, from Theorem 3.1 the above inequality condition $$|\bigcap_{i \in I}  {\bf S}_i| \leq n-d_{|I|}$$ is not sufficient for the existence of support constrained generator matrices.\\

\section{Conclusion}

The GM-MDS conjecture then theorem is a  beautiful result about the support configurations of codewords in generator matrices of  linear $1$-MDS codes. We show that the MDS condition in the GM-MDS conjecture is a special case of  a more generalized constraint based on the generalized Hamming weights for general linear codes. Then the direct generalization of the GM-MDS conjecture to $2$-MDS codes and algebraic geometric codes is proved not true over arbitrary finite fields.  Motivated from our GHW-based necessary condition for support constrained generator matrix, a GHW-based sufficient condition for support generator matrices for general linear codes is proved. Then a weaker sufficient condition for support constrained generator matrices of the the binary simplex code and the first order $q$-ary Reed-Muller codes is also presented.  \\

\end{document}